\documentclass[letterpaper, 10pt, conference]{ieeeconf}      

\IEEEoverridecommandlockouts                              
\usepackage{graphics} 
\usepackage{amssymb}  
\usepackage{mathtools}
\usepackage{hyperref}
\usepackage{cite}
\usepackage{multicol}
\usepackage{color}
\usepackage{url}

\newtheorem{theorem}{Theorem}

\newtheorem{definition}{Definition}
\newtheorem{proposition}{Proposition}

\newtheorem{remark}{Remark}

\title{\LARGE \bf
	A tunable mixed feedback oscillator}

\author{Weiming Che, Fulvio Forni
	\thanks{W. Che is supported by CSC Cambridge Scholarship. W. Che and F. Forni are with the Department of Engineering, University of Cambridge, CB2 1PZ, UK {\tt\small wc289|f.forni@eng.cam.ac.uk}}}

\begin{document}
	
	\maketitle
	\thispagestyle{empty}
	\pagestyle{empty}
	
	\begin{abstract}
		The interplay of positive and negative feedback loops on different time scales
		appears to be a fundamental mechanisms for robust and tunable oscillations in both biological systems and electro-mechanical systems. We develop a detailed analysis of a basic three dimensional Lure model to show how controlled oscillations arise from the tuning of positive and negative feedback strengths. Our analysis is based on dominance theory and confirms, from a system-theoretic perspective, that the mixed feedback is a fundamental enabler of robust oscillations. Our results
		are not limited to three dimensional systems and extend to larger systems via passivity theory, and to uncertain systems
		via small gain arguments.
	\end{abstract}

	\section{Introduction}
	Oscillations are central in biology and engineering. In biology, oscillators appear in several biochemical processes \cite{Goldbeter1997}, and play a crucial role in the sensory-information processing of neurons\cite{sherman2001tonic} and in rhythmic regulation in the body, such as breathing, chewing and walking \cite{marder2001central}. In engineering, the interests in oscillators dates back to the 19th century, with the development of oscillating electrical circuits to support applications like digital clock generators and audio/video  broadcast \cite{tucker1972history,Ginoux2012}. More recent efforts in 
	bio-inspired engineering aim at the design of tunable and resilient artificial oscillators for gait control in robotics \cite{ijspeert2007swimming,kimura1999realization} and for neuromorphic circuits \cite{ribar2019neuromodulation}. 
	In spite of the substantial literature on the topic, the generation of self-excited oscillations in complex systems and the
	analysis of their stability/robustness features remain challenging questions \cite{ijspeert2008central}.

	The presence of both positive and negative feedback loops appears to be a recurrent structure in various biological oscillations  \cite{smolen2001modeling,mitrophanov2008positive,tsai2008robust}. 
	Through dynamic conductances, neurons are extremely effective at combining positive and negative feedback, at several time scales, to achieve robust behavior and modulation capabilities in an uncertain setting  \cite{Marder2014,Drion2015b,sepulchre2019control}. 
	This is confirmed by the engineering design of electrical oscillators, where the combination of positive and negative feedback pervades the early development of electrical amplifiers \cite{tucker1972history,Bernstein2002}, and is at the core of classical relaxation oscillator circuits \cite{Chua1987,Ginoux2012}.
	
	The challenge is to achieve robust oscillations in spite of unreliable components and dynamical uncertainties. 
	Positive feedback increases the system sensitivity to input variations, leading to hysteresis, while negative feedback reduces this sensitivity, linearizing/normalizing the closed-loop behavior. Oscillations emerge from the interaction of these two feedback loops on different time scales \cite{sepulchre2019control}. The combination of fast positive feedback and slow negative feedback generates a non-minimal zero in open loop, which becomes a source of instability in closed loop that leads to robust controlled oscillations. 	
	The aim of this paper is to develop a system theoretic analysis of this important phenomena.
	
	Our analysis takes advantage of dominance theory and differential dissipativity
	\cite{forni2018differential,miranda2018analysis},
	rooted in the theory of monotone systems with respect to high rank
	cones \cite{Smith1980,Smith1986,Sanchez2009,Sanchez2010}.
	Dominance theory allows us to identify/enforce a low dimensional 
	``dominant" behavior (i.e. oscillations) in systems of any dimension,
	overcoming the usual restrictions to planar dynamics of Poincar{\'e}-Bendixson theory.
	It also allows us to analyze cases where describing function analysis shows limitations,
	typically due to dynamics that are not well represented by low harmonic approximations.
	Differential dissipativity  provides instead the framework to tackle interconnections. 
	
	We study the effects of combined positive and negative feedback in the simple setting of Lure system modeling. 
	In its simplest form, the controller is given by the parallel interconnection of two stable 
	first order linear networks, one acting with positive sign and the other with negative sign. 
	The positive branch is faster than the negative branch. Their relative strength is 
	regulated by a \emph{balance} parameter $\beta$
	while their collective strength is regulated by a \emph{gain} parameter $k$. 
	The controller acts on the plant, another stable first order linear network,
	through a simple sigmoidal saturation. The saturation is the only nonlinear element in our analysis.  
	We close the loop by feeding back the output of
	the plant to the controller, as summarized in
	Fig. \ref{fig:Block}. We call this closed loop \emph{mixed feedback amplifier}, to emphasize
	the crucial interplay of positive and negative feedback in determining the system behavior.
	
	In what follows we take advantage of root locus analysis and Nyquist analysis,
	adapted to dominance theory, to derive sufficient conditions for controlled oscillations.
	We show how the closed loop behavior can be modulated from a stable equilibrium into
	stable oscillations by tuning the balance and the gain of the feedback. Taking inspiration from
	conductance-based models in neuroscience, we then generalize the analysis to controllers based on several
	parallel feedback lines (positive and negative). We also derive results for feedback 
	interconnections, based on passivity,
	and we briefly touch upon the important topic of robustness to dynamic uncertainties.
	
	The paper is organized as follows. Section \ref{sec:model} presents the main model equations. 
	Section \ref{sec:dominance} is a brief introduction to dominance theory. 
	Section \ref{sec:dominance_feedback_tuning} shows the connection between the feedback
	parameters and the dominance properties of the closed loop. Section \ref{sec:fco} completes
	the design of the feedback parameters for oscillations. Extensions and interconnections are 
	discussed in Section \ref{sec:large_sys}. Conclusion follows. The proofs of the theorems can be found in the \href{https://arxiv.org/abs/2011.08564}{arXiv version} \cite{che2020tunable}.

	\section{The Mixed Feedback amplifier}
	\label{sec:model}
	
	We propose a simple mixed feedback amplifier model adapted from  \cite[Section III.C]{Drion2015b} and from
	the recently proposed neuromorphic circuit model \cite{ribar2019neuromodulation}, 
	which shows how oscillations can be shaped via system parameter tuning. 
	
\begin{figure}[!h]
		\centering
		\includegraphics[width=0.8\columnwidth]{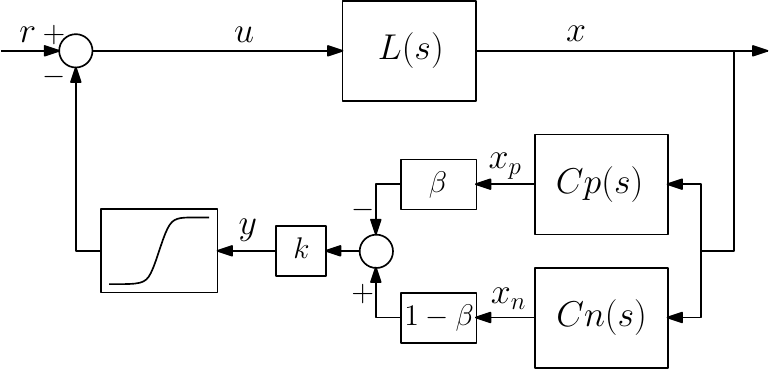}
		\caption{Block diagram of the mixed feedback amplifier.}
		\label{fig:Block}
	\end{figure}

	The mixed feedback amplifier consists of three parts: a load associated with state variable $x$ mimicking the passive membrane of a neuron, combined with positive feedback with state variable $x_p$, and negative feedback with state variable $x_n$. The feedback signals are summed and fed into a static sigmoidal saturation nonlinearity before closing the loop as shown in Fig. \ref{fig:Block}. Load, positive, and negative networks are simple first-order lags, a constraint that we will relax later in the paper.
	The resulting amplifier is represented by the differential equations
	\begin{equation}\label{sys:mixed_feedback}
		\left\{
		\begin{array}{rcl}
			\tau_l\dot{x} &=&-x+u\\
			\tau_p\dot{x}_p &=& x-x_{p}\\
			\tau_n\dot{x}_n &=& x-x_{n}\\
			u &=& -\varphi(y) + r \\
			y &=& k( - \beta x_p + (1-\beta)x_n)
		\end{array}
		\right.
	\end{equation}
	where $\tau_l, \tau_p, \tau_n$ are time constants of load,  positive, and negative feedback networks respectively; $r$ is the external input, typically held constant. The Lure structure of the closed loop is emphasized by the selection of the load input $u$ and of the mixed feedback output $y$. The mixed feedback is characterized by two main parameters, the \emph{gain} $k \geq 0$ and the \emph{balance} $0 \leq \beta \leq 1$, the latter capturing the relative strength between positive and negative feedback.
	
	The transfer functions of load, positive, and negative networks respectively read 
	\begin{equation}
		L(s)=\frac{1}{\tau_l s+1} \,,\ Cp(s)=\frac{1}{\tau_p s+1} \,,\ Cn(s)=\frac{1}{\tau_n s+1} \ .
	\end{equation}
	Each network is associated with a pole $p_l = - \frac{1}{\tau _l}$, $p_p = - \frac{1}{\tau _p}$ and $p_n = - \frac{1}{\tau _n}$. We assume a well-defined time scale separation,
	\begin{equation}
		\tau_p<\tau_n \ ,
	\end{equation} between the positive and negative feedback networks. 
	For simplicity of the exposition, we assume that 
	the time constant $\tau_l$ is never
	equal to $\tau_p$ or $\tau_n$. We also assume that
	$\varphi$ is slope-restricted, differentiable function
	\begin{equation}
		0 \leq \varphi' \leq 1
	\end{equation} 
	and such that {$\varphi'(y) = 0$ whenever $|y|$ is larger than a given threshold $M$
		($M$ can be any).}
	For simplicity, simulations  will use $\varphi=\tanh$.	
	
	We look at \eqref{sys:mixed_feedback} as a Lure system, which is the negative feedback interconnection of the transfer function from input $u$ to output $y$ given by
	\begin{equation}\label{eq:linear_tf}
		G(s,k,\beta)=\frac{-k\Big(\big(\beta(\tau_n+\tau_p)-\tau_p\big)s+2\beta-1\Big)}{(\tau_ls+1)(\tau_ps+1)(\tau_ns+1)} 
	\end{equation}
	and the sigmoidal nonlinearity $\varphi$. This allows us to examine the effects of variations in the mixed feedback gain and balance by using classical linear tools, like root locus analysis, Nyquist diagrams, and a generalized circle criterion adapted to the recently developed dominance theory \cite{miranda2018analysis,forni2018differential}.
	
	\section{Dominance Theory} 
	\label{sec:dominance}
	
	Dominance theory aims to capture the existence of low dimensional attractors in a high dimensional system state space.
	For linear system $\dot{x}=Ax$, $x \in \mathbb{R}^n$, $p$-dominance means that the dynamics can be split into  $p$ slow dominant modes and $n-p$ fast transient modes. For nonlinear systems
	\begin{equation} \label{eq:general_nonlinear}
		\dot{x}=f(x) \, , \quad x\in\mathbb{R}^n
	\end{equation}
	dominance can be characterized by using the linearization along the {system trajectories 
		(capturing infinitesimal variations in the neighborhood a generic trajectory $x(\cdot)$)}
	\begin{equation}\label{eq:prolonged_sys}
		\begin{cases}
			\dot{x}=f(x)\\
			\delta \dot{x}=\partial f(x) \delta x
		\end{cases}\quad (x,\delta x)\in \mathbb{R}^n\times\mathbb{R}^n \ .
	\end{equation} 
	\begin{definition}
		The nonlinear system \eqref{eq:general_nonlinear} is $p$-dominant with rate $\lambda\geq0$ if and only if there exist a symmetric matrix P with inertia $(p,0,n-p)$ and $\varepsilon\geq0$ {such that \eqref{eq:prolonged_sys} satisfies the} conic constraint
		\begin{equation}
			\begin{bmatrix}
				\delta\dot{x}\\
				\delta x
			\end{bmatrix}^T \begin{bmatrix}
				0&P\\P&2\lambda P+\varepsilon I
			\end{bmatrix}\begin{bmatrix}
				\delta\dot{x}\\
				\delta x
			\end{bmatrix}\leq 0
		\end{equation}
		along all its trajectories. The property is strict if $\varepsilon>0$.$\hfill\lrcorner$
	\end{definition}
	Here the symmetric matrix $P$ with inertia $(p,0,n-p)$ has $n-p$ positive eigenvalues and $p$ negative eigenvalues;
	$p$ is the dimension of the dominant sub-dynamics of the nonlinear system. We are particularly interested in small $p$ as this entails that the nonlinear system possesses a simple attractor, {as summarized in the next theorem, \cite[Corollary 1]{forni2018differential}.}
	\begin{theorem}\label{th:p-attractor}
		For a strict $p$-dominant system \eqref{eq:general_nonlinear} with dominant rate $\lambda \ge 0$, every bounded trajectory asymptotically converges to
		\begin{itemize}
			\item a unique fixed point if $p=0$;
			\item a fixed point if $p=1$;
			\item a simple attractor if $p=2$, that is, a fixed point, a set of
			fixed points and connecting arcs, or a limit cycle.$\hfill\lrcorner$
		\end{itemize} 
	\end{theorem}
	For the analysis of the mixed feedback amplifier we are particular interested in the case $p=2$, where the asymptotic behavior of the system is essentially captured by a planar system. As a result, we can predict the existence of the stable oscillations if all equilibrium points of the system are unstable and trajectories remain bounded.
	\begin{figure}[!h]
		\centering
		\includegraphics[width=0.55\columnwidth]{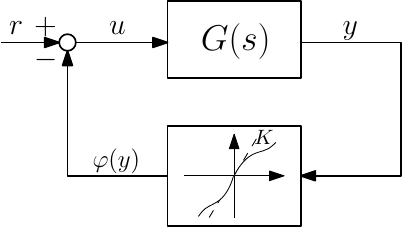}
		\caption{Lure feedback system}
		\label{fig:Lure_theory}
	\end{figure}
	
	For Lure feedback systems, $p$-dominance can be also characterized in the frequency domain, through the circle criteria for $p$-dominance, \cite[Corollary 4.5]{miranda2018analysis}.
	\begin{theorem}[Circle criteria for $p$-dominance]\label{th:circle_cirteria}
		Consider the Lure feedback system in Fig. \ref{fig:Lure_theory} given by the negative feedback interconnection of the linear system $G(s)$ and the static nonlinearity $\varphi$ satisfying sector condition $\partial\varphi\in(0,K)$. The closed system is strictly $p$-dominant with rate $\lambda$ if:
		\begin{enumerate}
			\item the real parts of all the poles of $G(s)$ not equal to $-\lambda$;
			\item the shifted transfer function $G(s-\lambda)$ has $p$ unstable poles;
			\item the Nyquist plot of $G(s-\lambda)$ lies to the right hand side of the line $-1/K$ on the Nyquist plane. $\hfill\lrcorner$
		\end{enumerate}
	\end{theorem}
	
	Theorem \ref{th:circle_cirteria} will be extensively used in this paper.
	It will be particularly useful when combined to the notion of $p$-\emph{passivity}, \cite{forni2018differential}. 
	Indeed, a linear system with transfer function $G(s)$ is $p$-passive with rate $\lambda$ if 
	it satisfies the conditions of Theorem \ref{th:circle_cirteria} for $K=\infty$. In such a case,
	the Lure system in Fig. \ref{fig:Lure_theory} is also $p$-passive from $r$ to $y$, \cite{miranda2018analysis}.
	Thus, a $p$-passive system is a $p$-dominant system. Furthermore,  $p$-passivity
	is preserved by negative feedback interconnections, as clarified by the following theorem, 
	\cite{forni2018differential}.
	\begin{theorem}\label{thm:passive_interconnection}
		Consider two systems $\Sigma_i$ and assume they are respectively $p_i$-passive with common rate $\lambda$ with input $u_i$ and output $y_i$, where $i \in \{1,2\}$. Then, the negative feedback loop given by $u_1 = -y_2 + r_1$ and $u_2 = y_1 + r_2$ is $(p_1+p_2)$-passive from $(r_1,r_2)$ to $(y_1,y_2)$.
	\end{theorem}
	
	
	
	\section{Feedback tuning via dominance theory}
	\label{sec:dominance_feedback_tuning}
	
	The linear system $G(s,k,\beta)$ in \eqref{eq:linear_tf} is stable with poles in $-\frac{1}{\tau_l}$, $-\frac{1}{\tau_p}$, and $-\frac{1}{\tau_n}$. Once the time constants are fixed, the system's behavior is modulated by the balance $\beta$ and by the gain $k$. $\beta$ determines the position of the zero of $G(s,k,\beta)$, 
	\begin{equation} \label{eq:main_zero}
		z=\frac{2\beta-1}{\beta(\tau_p+\tau_n)-\tau_p} \ .
	\end{equation}
	The zero has positive real part for $\beta > 0.5$, {i.e. when} the contribution of the 
	positive feedback component is sufficiently strong.
	Combined with a large gain $k$, looking at the root locus of the linearized closed loop, the zero acts as an ``attractor" for the closed-loop poles, to guarantee (a controlled) instability of the unsaturated closed loop equilibrium (for $r =0$). This zero and its modulation via $\beta$ is the fingerprint of the mixed feedback amplifier. It {guarantees robust} modulation between the stable regime and the unstable/oscillatory regime of the closed-loop system.
	
	{For all $\beta\in[0,1]$, $G(s,k,\beta)$ is a stable} transfer function therefore
	the closed loop remains stable for sufficiently small gain $k$, as certified by the following theorem.
	\begin{theorem}\label{th:0-dominance}
		For any constant $r$ and any $\beta\in[0,1]$, the system \eqref{sys:mixed_feedback} is 0-dominant with $\lambda=0$ 
		for any  gain $0 \leq k < \overline{k}_0 $ given by
		\begin{equation} \label{eq:k0}
			\overline{k}_0 = 
			\begin{cases}
				\infty \! & \! \mbox{if } \min\nolimits\limits_{\omega} \Re(G(j \omega ,\!1,\!\beta)) \!\geq\! 0 \\
				- \dfrac{1}{\min\nolimits\limits_\omega \Re(G(j\omega ,\!1,\!\beta))} \!&\!  \mbox{otherwise. } 	
			\end{cases} \vspace{-1mm}
		\end{equation}
		$\hfill\lrcorner$
	\end{theorem}
	
\begin{figure}[htbp]
		\centering
		\includegraphics[width=0.55\columnwidth]{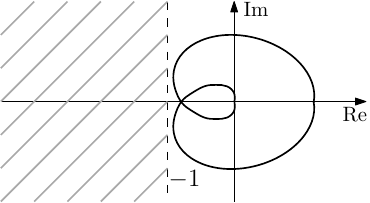}
		\caption{An illustration of the proof of $0$-dominance using circle criteria. The shaded region denotes the constraint to the Nyquist plot of $G(s,k,\beta)$, set by nonlinearity $\varphi$. The Nyquist locus represents the case of a zero in \eqref{eq:main_zero} has positive real part.}
		\label{fig:circle_criteria_zero}
	\end{figure}

	Using the circle criterion, the closed system is $0$-dominant as long as the Nyquist plot of \eqref{eq:linear_tf} remains to the right of the shaded region in Fig. \ref{fig:circle_criteria_zero}. Given the boundedness of $|G(s,1,\beta)|_\infty$, the exact value of $\overline{k_0}$ can be determined numerically as in \eqref{eq:k0}.
	
	We now look into the tuning of $k$ and $\beta$  to enable oscillations, using the
	shifted transfer function $G(s-\lambda,k,\beta)$ and focusing on $2$-dominance. 
	As first step, the rate $\lambda$ must be chosen to ensure that $G(s-\lambda,k,\beta)$ has two unstable poles,
	a necessary condition for dominance due to {specific (differential) sector} condition $0 \leq \varphi' \leq 1$.
	\begin{theorem}\label{th:2-dominance}
		Consider a rate $\lambda$ for which the 
		transfer function $G(s-\lambda,k,\beta)$ has two unstable poles. Then, 
		for any constant $r$ and any $\beta\in[0,1]$, 
		the system \eqref{sys:mixed_feedback} is 2-dominant with rate $\lambda$ 
		for any  gain $0 \leq k < \overline{k}_2 $ given by
		\begin{equation} \label{eq:k2}
			\overline{k}_2 = 
			\begin{cases}
				\infty \! & \! \mbox{if } \min\nolimits\limits_{\omega} \Re(G(j \omega-\lambda ,\!1,\!\beta)) \!\geq\! 0 \\
				- \frac{1}{ \min\nolimits\limits_\omega \Re(G(j\omega-\lambda ,\!1,\!\beta))} \!&\!  \mbox{otherwise. } 	
			\end{cases}
		\end{equation}
	\end{theorem}
	The proof argument simply follows the one of Theorem \ref{th:0-dominance} applied to the shifted transfer function.
	
	Theorem \ref{th:0-dominance} and \ref{th:2-dominance} {show} that there is a range of gains $k$ for which 
	the system is at the same time $0$-dominant and $2$-dominant. This is not a contradiction, since the range 
	of behaviors of a $0$-dominant system is compatible with the range of behaviors of a $2$-dominant system. 
	The system behavior will be constrained by the most restrictive of the two properties, namely $0$-dominance, 
	ensuring that all trajectories converge to a unique equilibrium. As a result, in tuning parameters for oscillations, 
	we will explore the range of gains $\bar{k}_0 < k < \overline{k}_2$. 
	
	When the positive feedback is strong enough the close system \eqref{sys:mixed_feedback} becomes 
	$2$-passive, as clarified below.
	\begin{theorem}\label{th:2_passive}
		Consider the critical balance 
		$\beta^*=\frac{\tau_p}{\tau_p+\tau_n}$.
		For any $\beta\in(\beta^*,1]$, there exist a rate $\lambda \geq 0$ for which
		the system \eqref{sys:mixed_feedback} is $2$-passive from $r$ to $y$, for any $k \geq 0$. $\hfill\lrcorner$
	\end{theorem}
	
	{For a suitable selection of $\lambda$ and $\beta\in(\beta^*,1]$, 
		Theorem \ref{th:2_passive} guarantees that $\bar{k}_2 = \infty$ (in Theorem \ref{th:2-dominance}).
		We also observe that the critical balance $\beta^*$ 
		tends towards zero as the positive feedback component becomes faster, i.e. $\tau_p \to 0$.}
	Finally, taking advantage of Theorem \ref{thm:passive_interconnection}, $2$-passivity of \eqref{sys:mixed_feedback}
	means that we can interconnect the mixed feedback amplifier with other $0$-passive systems
	while preserving $2$-passivity.   
	
	\section{Feedback control of oscillations}
	\label{sec:fco}
	Based on the analysis in section IV, the parametric range for the mixed feedback amplifier to be $0$-dominant, $2$-dominant, and $2$-passive can be derived numerically for any given set of the time constants $\tau_l$, $\tau_p$, $\tau_n$ that satisfy the key assumption of time scale separation between positive and negative feedbacks. Oscillations will arise
	for gains $\bar{k}_0 < k < \overline{k}_2$, when the system is $2$-dominant but not $0$-dominant, as typically
	identified by the presence of (at least) one unstable equilibrium in closed loop. {The existence of 
		this unstable equilibrium is guaranteed by the attraction exerted on the closed loop poles by the unstable zero 		\eqref{eq:main_zero}, for a suitable selection of $\beta$ and for a sufficiently large gain $k$.
		In fact, the stability of the open-loop transfer function $G(s,k,\beta)$ combined with 
		the the conditions on $\varphi$ guarantee that the trajectories
		of the closed-loop system} are bounded for any balance $\beta\in[0,1]$ and {gain $k \geq 0$.} Therefore, by
	Theorem \ref{th:p-attractor}, oscillations will arise when the closed loop is $2$-dominant and has only unstable 
	equilibria. When some of the equilibria are stable, oscillations may coexist with them.
	
	\begin{figure}[b]
		\centering
		\includegraphics[width=0.9\columnwidth]{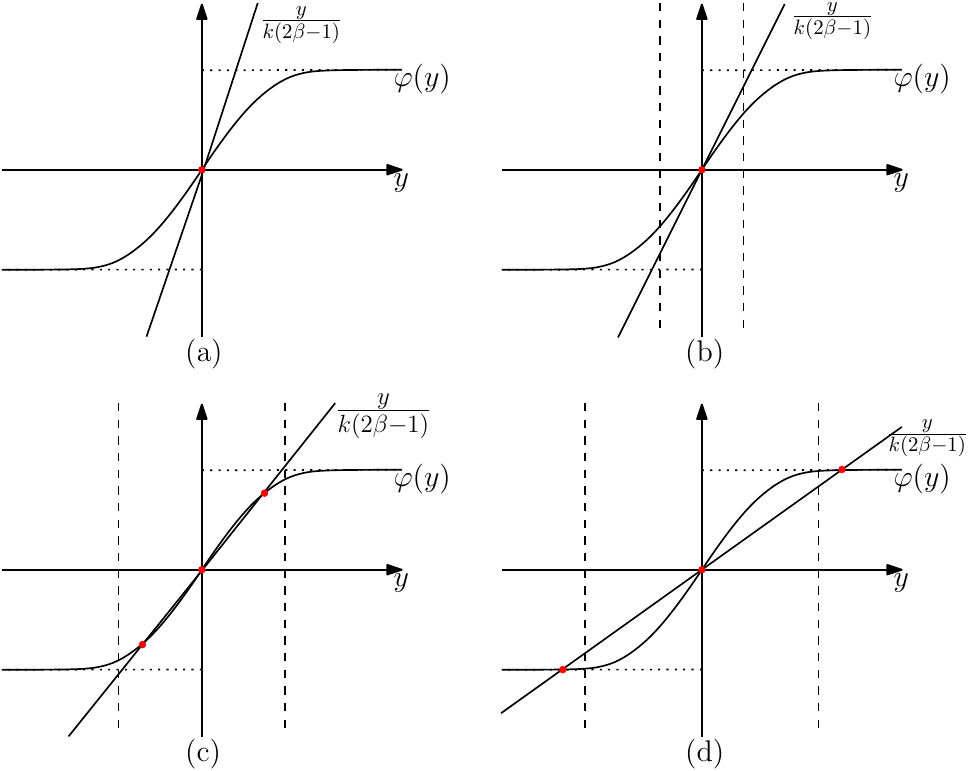}
		\caption{Four situations of the stability of the equilibrium points for $\beta\in(0.5,1)$, input $r=0$. The fixed points are unstable if it falls in the region bounded by the dashed lines. The case $r\neq0$ is similar, taking into account a vertical shift to $\varphi(y)$.}
		\label{fig:fixed_point_stability}
	\end{figure}
	
	The number of equilibria of the closed loop can be determined by looking at the intersection of the static nonlinearity $ \varphi(y)$ (here we use $\varphi=\tanh$) with the line $ \frac{y}{k(2\beta-1)}$, where $k(2\beta-1)$ is the DC gain of $G(s,k,\beta)$ \eqref{eq:linear_tf}. For $\varphi = \tanh$, when $\beta\in[0,0.5]$, we have that $k(2 \beta-1)<0$ therefore we have one equilibrium in zero. The equilibrium is stable if $k$ is smaller than the gain margin of $G(s,1,\beta)$. When $\beta\in(0.5,1]$, for small $k$ in Fig. \ref{fig:fixed_point_stability}a,   the lack of dashed lines shows that the equilibrium is stable and there are no oscillations. As $k$ and $\beta$ increases, the $0$ equilibrium becomes unstable and oscillations appear, Fig. \ref{fig:fixed_point_stability}b.
		For larger values, two (and more) unstable equilibria emerge, preserving oscillations, Fig. \ref{fig:fixed_point_stability}c.
		They eventually become stable, and  oscillations may disappear or coexist with stable equilibria,
		Fig. \ref{fig:fixed_point_stability}d. Exact gain $k$ and balance $\beta$ at which such stability change takes place can be determined numerically.

	\begin{figure}[t]
		\vspace*{-0.15in}
		\centering
		\includegraphics[width=1\columnwidth]{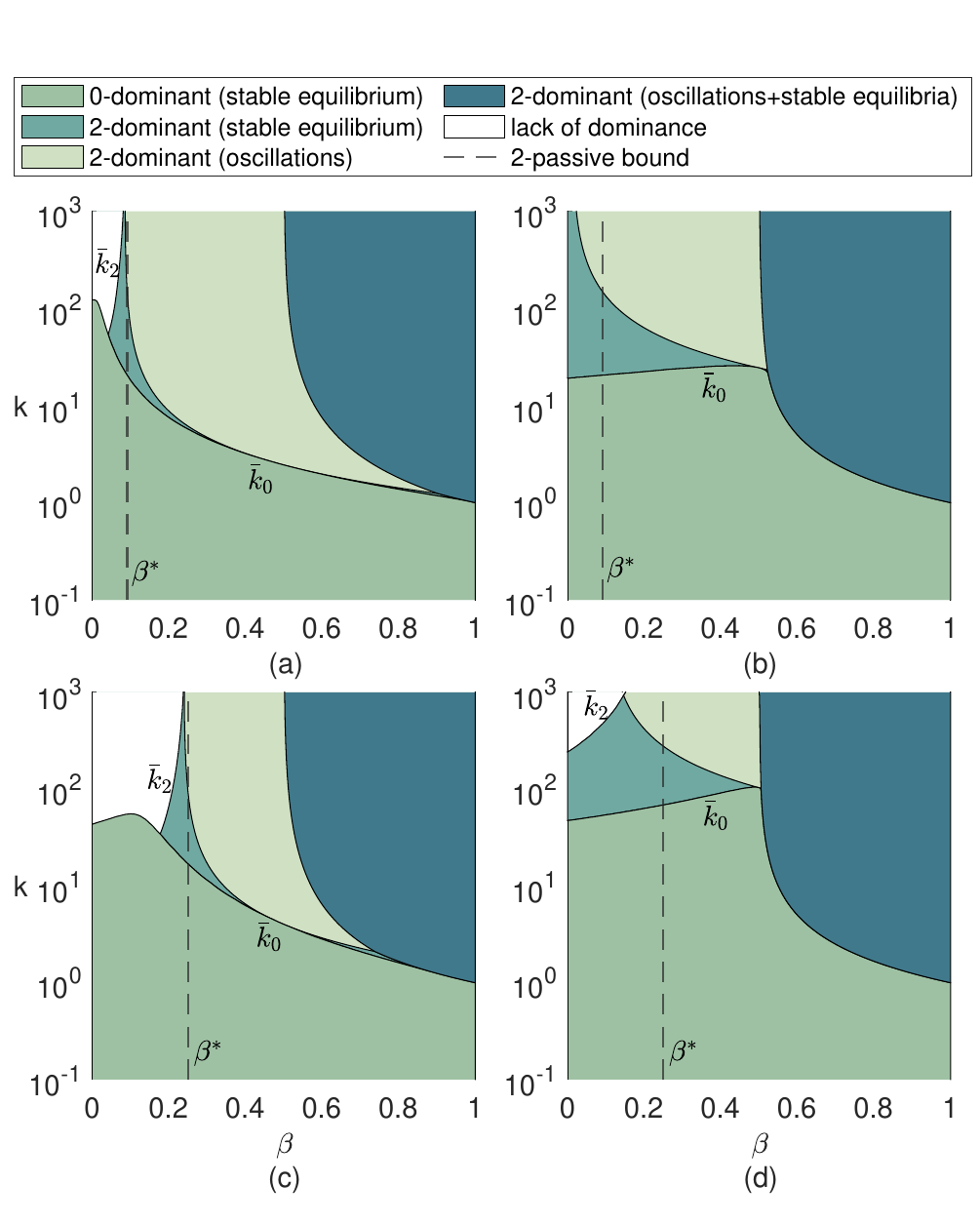}
		\vspace*{-0.15in}
		\caption{Dominance map for input $r=0$. \textbf{(a)} fast load: $\tau_l=0.01$, $\tau_p=0.1$, $\tau_n=1$, $\lambda=50$; \textbf{(b)} slow load: $\tau_l=10$, $\tau_p=0.1$, $\tau_n=1$, $\lambda=5$; \textbf{(c)} fast load, reduced time-scale separation: $\tau_l=0.01$, $\tau_p=0.1$, $\tau_n=0.3$, $\lambda=50$; \textbf{(d)} slow load, reduced time-scale separation: $\tau_l=10$, $\tau_p=0.1$, $\tau_n=0.3$ $\lambda=5$.}
		\label{fig:dominance_map}
	\end{figure}
	
	For illustration, we fix the reference $r=0$ and we derive four parametric maps on the dominance of the mixed feedback amplifier for $k\in(0.1,1000)$ and $\beta \in [0,1]$, for different time scale arrangements, as shown in Fig. \ref{fig:dominance_map}. Here, for simplicity, we set $\lambda$ roughly in the middle of the left-most pair of poles (this guarantees the additional property of 2-passivity).
	
	Moving horizontally on Fig. \ref{fig:dominance_map}, we see that increase in $\beta$ modulates the closed-loop behavior from global stability of an equilibrium into oscillations (eventually leading to multi-stability), for $k$ within a suitable range. Likewise, moving vertically on Fig. \ref{fig:dominance_map}, we see that oscillations are always triggered (dampened) by increasing (decreasing) the feedback gain $k$, for $\beta$ within a suitable range. Fig. \ref{fig:dominance_map} also shows numerically how a larger time scale separation guarantees a larger region of oscillations and a lower critical balance $\beta^*$ for $2$-passivity.
	
	\begin{figure}[htbp]
		\centering
		\vspace*{-0.15in}
		\includegraphics[width=1\columnwidth]{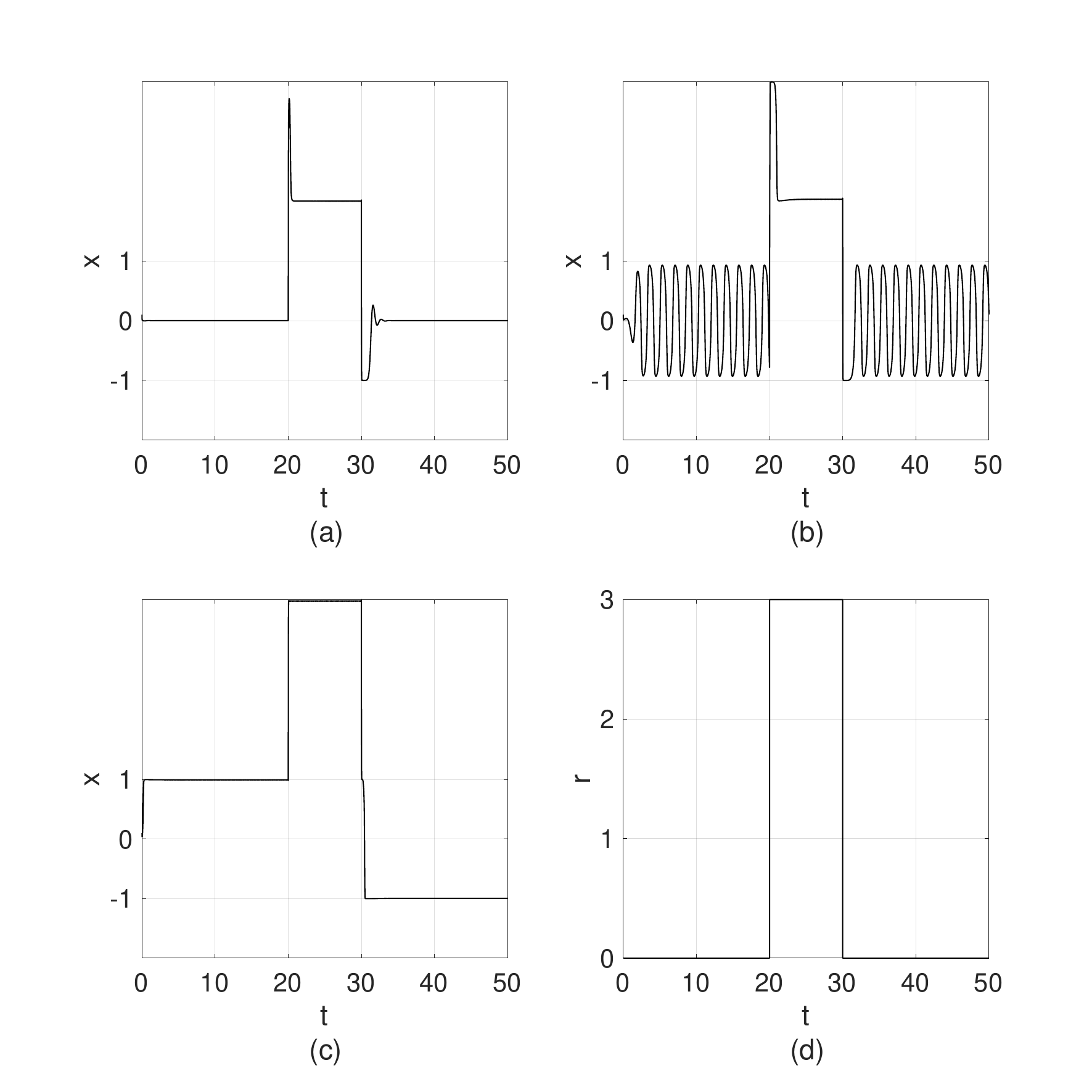}
		\vspace*{-0.15in}
		\caption{Simulations of the mixed-feedback amplifier in response to the input reference $r$,
			from the initial condition $x=0.1$,  $x_p=0$ and $x_n=0$, and different tuning of $k$ and $\beta$. 
			\textbf{(a)}: $k=5$, $\beta=0.2$ ($0$-dominant); \textbf{(b)}: $k=5$, $\beta=0.4$ ($2$-dominant,  oscillation); 
			\textbf{(c)}: $k=5$, $\beta=0.8$ ($2$-dominant oscillation + stable equilibria).}
		\label{fig:simu}
	\end{figure}
	
	We complete our discussion with simulations, in Fig. \ref{fig:simu}.
	We consider time constants as in Fig. \ref{fig:dominance_map}a
	and we select gain and balance to explore the three main regions
	in the figure: $0$-dominance is illustrated in Fig. \ref{fig:simu}a,
	$2$-dominance (oscillations) is in Fig. \ref{fig:simu}b, and the
	last case of $2$-dominance (oscillations + stable equilibria) is in 
	Fig. \ref{fig:simu}c. Fig. \ref{fig:simu}d illustrates the piecewise constant input $r$
	used to {perturb} the system from its steady state for $20\leq t\leq30$ (bottom right). 
	After the step change in the reference $r$, the state trajectories of the system tuned in the $0$-dominant region,
	Fig. \ref{fig:simu}a, and in the $2$-dominant (oscillation) region, Fig. \ref{fig:simu}b, converge back to the original steady state. For the system tuned in the $2$-dominant (oscillations + stable equilibria) region, Fig. \ref{fig:simu}c, 
	the trajectory starts from one of the stable equilibria and, after reference perturbation, converges to the other one.

	\section{LARGE MIXED FEEDBACK SYSTEMS}
	\label{sec:large_sys}
	
	\subsection{Multi-channel mixed feedback amplifier}
	The results presented in the previous sections can be extended to larger systems.
	As first example, in this section we take inspiration from neuroscience,
	where the membrane electrical potential of a neuron is modulated by 
	multiple parallel ion channels \cite{hodgkin1952quantitative,Marder2014,Drion2015b}. Thus, we show how the mixed feedback structure 
	can be extended to a larger parallel of first order systems 
	while preserving the overall closed-loop behavior and modulation features.
	In fact, looking at  Fig. \ref{fig:Block}, for a slow first order $L(s)$ and 
	for $Cp(s)$ and $Cn(s)$ given by the sum of several first order networks with similar time scales, 
	$\sum_{i}\frac{1}{\tau_i s - 1}$, the resulting system
	shows clustered and interlaced poles and zeros, which have negligible effects 
	on the dominance properties to the system.
	
	\begin{proposition}\label{th:Interlacing2}
		Let $C(s)$ be the difference of two finite parallels of first order transfer functions
		separated by a well-defined time scale:
		\begin{equation}\label{eq:Interlacing_expression}
			\begin{split}
				C(s)&=\beta Cp(s)- (1-\beta)Cn(s)\\
				&=\beta\sum_{i=1}^{m}\frac{\rho_i}{\tau_is+1}-(1-\beta)\sum_{i=m+1}^{m+n}\frac{\rho_i}{\tau_is+1} \ ,
			\end{split}
		\end{equation}
		where 
		\[-\frac{1}{\tau_1}<-\frac{1}{\tau_2}<....<-\frac{1}{\tau_m}< -\frac{1}{\tau_{m+1}}<...<-\frac{1}{\tau_{m+n}}\]  and 
		$\sum_{i=1}^{n}\rho_i=1$ and $\sum_{i=m+1}^{m+n}\rho_i=1$ (unit gain). Then for $\beta\in(0,1),$
		\begin{itemize}
			\item the poles of $C(s)$ are those of $Cp(s)$ and $Cn(s)$.
			\item $C(s)$ has $m+n-1$ zeros, where\begin{itemize}
				\item $m-1$ zeros are interlaced with the poles of $Cp(s)$ on real axis;
				\item $n-1$ zeros are interlaced with the poles of $Cn(s)$ on real axis;
				\item one zero lies on the real axis and locates in $(-\infty,-1/\tau_1)\cup(-1/\tau_{m+n},\infty)$. $\hfill\lrcorner$
			\end{itemize}
		\end{itemize}
	\end{proposition}
	
\begin{figure}[!h]
		\centering
		\includegraphics[width=0.8\columnwidth]{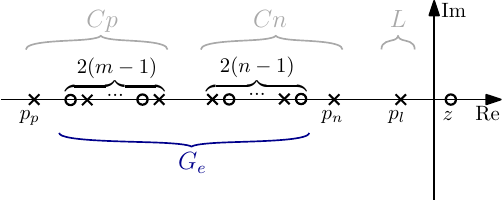}
		\caption{An illustration of pole-zero distribution of $\overline{G}(s,k,\beta)$.}
		\label{fig:dominance_InterPZ}
	\end{figure}

	Proposition \ref{th:Interlacing2} makes straightforward to generalize dominance analysis
	to the high dimensional mixed feedback amplifier. The extended open-loop transfer 
	function $\overline{G}(s,k,\beta)$ has the structure
	\begin{equation}\label{eq:Interlacing_sys_TF}
		\overline{G}(s,k,\beta)=G(s,k,\beta)G_e(s)
	\end{equation}
	where $G(s,k,\beta)$ corresponds to \eqref{eq:linear_tf},
	with a fast pole $p_p$, two slow poles $p_n$ and $p_l$ and a right-most zero $z$
	(for sufficiently large $\beta$), 
	and $G_e(s)$ is a bi-proper transfer function collecting the remaining poles and zeros, as shown 
	in Fig. \ref{fig:dominance_InterPZ}).

	Under the assumptions of Proposition  \ref{th:Interlacing2},
	it is straightforward to show that 
	Theorems \ref{th:0-dominance} and \ref{th:2-dominance} hold for the Lure system
	given by $\overline{G}(s,k,\beta)$ and $\varphi$ (using the same proof argument).
	Furthermore, in spite of the extended system dimension, the number and stability of 
	the closed-loop equilibria can be analyzed as in Section \ref{sec:fco}, looking at the intersection between
	$\varphi$ and the line with slope given by $\overline{G}(j0,k,\beta)$, 
	and to the root locus of $\overline{G}(s,k,\beta)$. Thus, through the root locus, their 
	stability is fundamentally related to the relative degree 
	of $\overline{G}(s,k,\beta)$ combined to the presence of an unstable right-most zero 
	($z$ in figure \ref{fig:dominance_InterPZ}). This means that Fig. \ref{fig:fixed_point_stability}
	holds thus, for sufficiently large $\beta$ and $k$, the system will oscillate as the consequence
	of $2$-dominance combined with unstable equilibria.

	\subsection{Passive external loads}
\begin{figure}[!h]
		\centering
		\includegraphics[width=0.6\columnwidth]{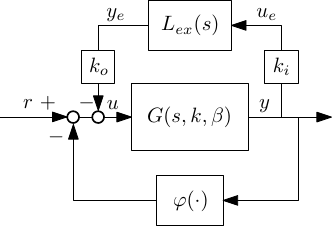}
		\caption{The interconnection of load and the mixed feedback amplifier}
		\label{fig:Load_block}
	\end{figure}

	Theorem \ref{th:2_passive} shows that the mixed feedback amplifier \eqref{sys:mixed_feedback}
	is $2$-passive from $r$ to $y$ for a suitable selection of balance $\beta$ and gain $k$. Thus,
	by Theorem \ref{thm:passive_interconnection}, feedback interconnections with any 
	$0$-passive system (sharing the same rate $\lambda$) will preserve $2$-passivity, i.e. $2$-dominance. 
	This suggests a route to build larger 2-passive systems through interconnection,
	as shown in Fig. \ref{fig:Load_block}. 
	It also illustrates the role of the mixed feedback amplifier as a feedback controller 
	driving stable $0$-passive loads into oscillations. 
	
	We consider here a linear load $L_{ex}(s)$, modeling a generic  
	mass-spring-damper system. The interface between mixed feedback amplifier
	and external load is represented by simple linear coefficients $k_o>0$ and $k_i>0$, as shown in in Fig. \ref{fig:Load_block}. 
	The mass-spring-damper system transfer function from the force input $u_e$ to the mixed velocity/position
	output $y_e$ reads
	\begin{equation}\label{Load:mechancial_load}
		L_{ex}(s)=\frac{k_vs+k_p}{s^2+bs+a}
	\end{equation}
	where $b>0$ and $a>0$ are normalized damping coefficient and spring constant, respectively.
	The output is a weighted combination of velocity and position variables,
	through the gains $k_v>0$ and $k_p>0$. 
	
\begin{figure}[!h]
		\centering
		\includegraphics[width=0.95\columnwidth]{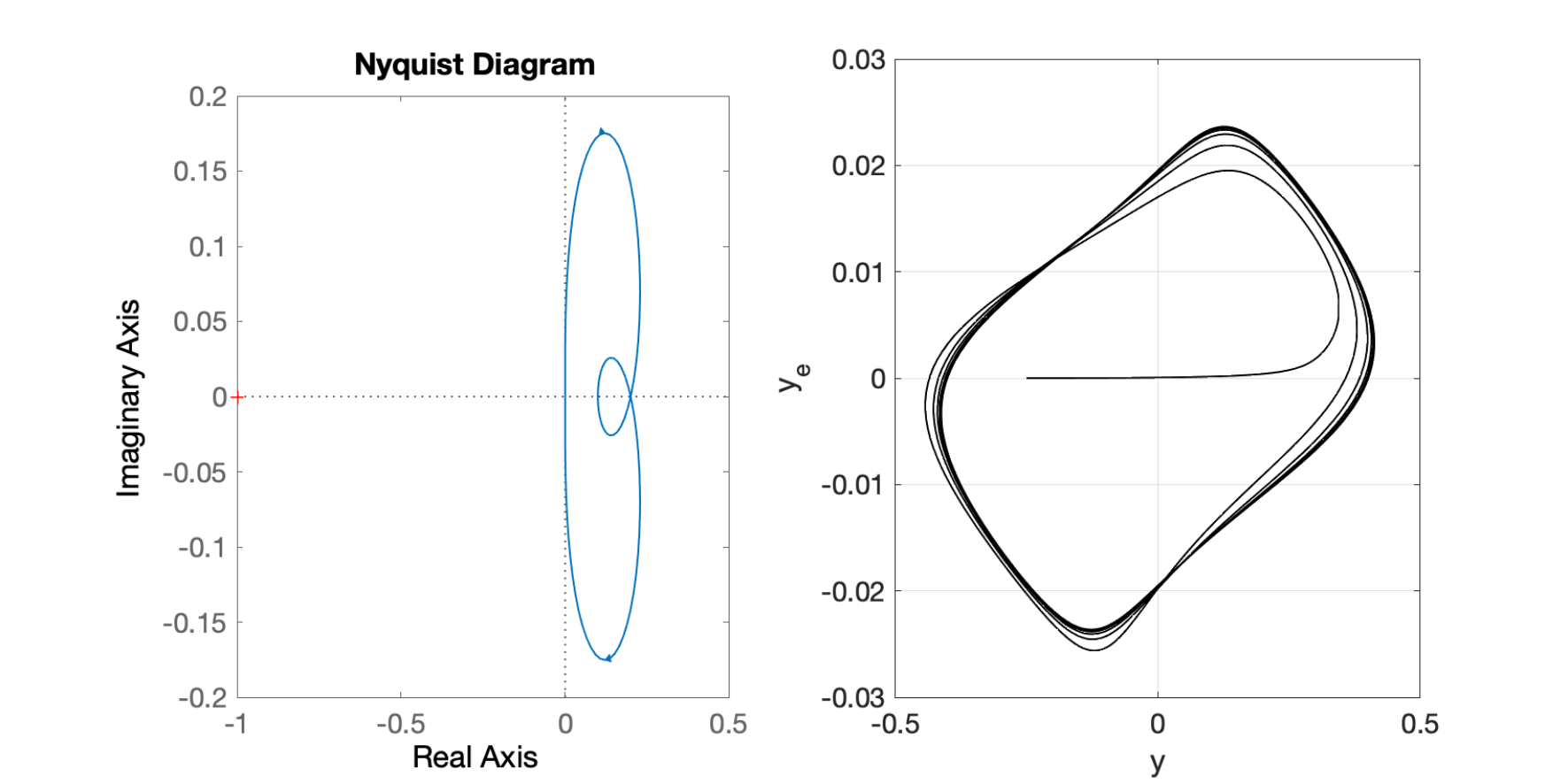}
		\caption{Left: Shifted Nyquist plot of the mass-spring-damper system $L_{ex}(s-\lambda)$; $\lambda=15,\ k_v=1,\ k_p=20,\ d=35,\ k=350$; Right: Trajectory of interconnected system projected on $(y,y_e)$ plane, with $\tau_l=0.01$, $\tau_p=0.1$, $\tau_n=1$, $k=10$, $\beta=0.4$,  $k_{i}=10$, $k_{o}=1$.}
		\label{fig:Load_example}
	\end{figure}

	To build a $2$-passive interconnection, 
	we need to show $0$-passivity of the the mass-spring-damper
	system for a given rate $\lambda \geq 0$.
	For the sake of illustration, consider
	$b =35$ and $a=350$ which give stable poles with real part smaller than $-\lambda$, 
	for $\lambda = 15$. For these parameters, $L_{ex}(s)$ is $0$-passive for $k_v = 1$ and $k_p = 20$,
	which guarantee that $L_{ex}(s-\lambda)$ is positive real as illustrated by the Nyquist plot
	in Fig. \ref{fig:Load_example} (left). Note that, differently from the case $\lambda = 0$,
	velocity feedback is not enough to guarantee $0$-passivity
	for $\lambda > 0$. A combination of velocity and position feedback is required to enforce $0$-passivity, 
	through the phase contribution of a stable zero in the transfer function. 
	
	Given the parameters of the mass-spring-damper system, 
	we consider a mixed feedback amplifier based on a fast internal load $L(s)$
	with time constant $\tau_l = 0.01$ (negligible), and positive and negative feedback time constants 
	$\tau_p = 0.1$ and $\tau_n =1$. The mixed feedback amplifier is tuned for $2$-passivity with rate $\lambda = 15$
	by taking gain $k=10$ and balance $\beta =0.4$, which also correspond to an oscillatory regime for $r=0$
	for the mixed feedback amplifier in isolation.
	
	From Theorem \ref{thm:passive_interconnection}, the negative feedback interconnection of the mixed 
	feedback amplifier with the mass-spring-damper system is $2$-passive. 
	Preserving $2$-passivity does not guarantee that the feedback loop
	will continue to oscillate, since $2$-passivity is also compatible with the presence of a stable equilibrium.
	However, if the overall closed-loop equilibria are unstable and trajectories are bounded, the closed-loop system
	will oscillate. In the example this is achieved by tuning the 
	interconnection gains $k_i$ and $k_o$. A stable oscillatory regime is obtained for 
	$k_o = 1$ and $k_i = 10$, as shown in Fig. \ref{fig:Load_example} (right).
	
	\begin{remark}  
		In comparison to entrainment phenomena, which are typically identified by a cascade structure, 
		these oscillations arise from the feedback interconnection of the mixed feedback amplifier 
		with an external load. They are thus shaped by the respective features of the two systems. 
		This means, for example, that oscillations may emerge from the interconnection
		of the mixed feedback amplifier with an external load even if the two systems
		in isolation do not oscillate.
	\end{remark}
	
	\subsection{Robustness to unmodeled dynamics}
	
\begin{figure}[!h]
		\centering
		\includegraphics[width=0.9\columnwidth]{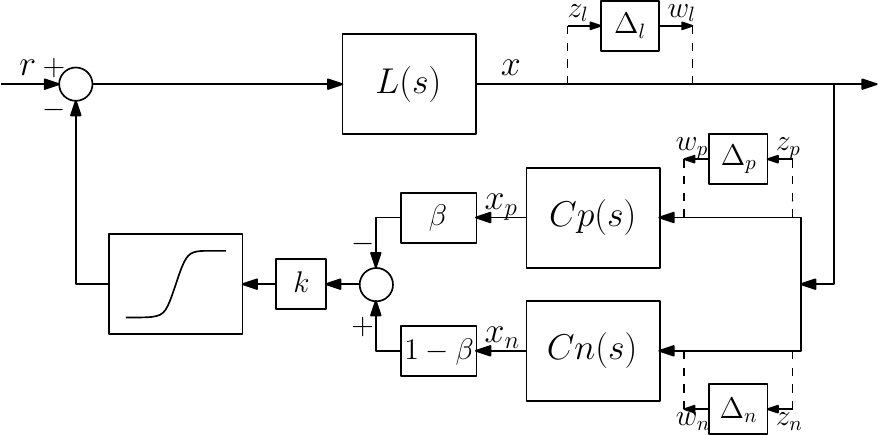}
		\caption{The mixed feedback amplifier with multiplicative unmodeled dynamics $\Delta_l$, $\Delta_p$, and $\Delta_n$ associated with $L(s)$, $Cp(s)$, and $Cn(s)$.}
		\label{fig:uncertainty_block}
		\vspace*{-0.2cm}
	\end{figure}

	We briefly touch upon the robustness properties of the mixed feedback amplifier,
	which can be 
	discussed within the classical framework of robust control, using dynamic
	uncertainties $\Delta_l$, $\Delta_p$, $\Delta_n$ as shown in Fig. \ref{fig:uncertainty_block}.
	
	A qualitative perspective is provided by Theorem \ref{th:circle_cirteria},
	which makes clear that small perturbations on the shape of 
	the Nyquist plot of the open-loop transfer function $G(s,k,\beta)$ cannot change the $p$-dominance of the system.
	This simple graphical observation suggests that $p$-dominance is a robust property.
	For a quantitative perspective, 
	the robustness of $p$-dominance property of the mixed-feedback amplifier can be studied 
	using small-gain results, adapted to dominance through the notion of $p$-gain. 
	Not surprisingly, the feedback interconnection of a $p$-dominant system with $0$-dominant 
	uncertainties $\Delta_l$, $\Delta_p$, and $\Delta_n$ remains $p$-dominant if the product of their gains 
	is sufficiently small (their product must be less than one, as
	in the classical stability \cite{forni2018differential,padoan2019h}). 
	
	In summary, small $0$-dominant perturbations with rate $\lambda$, 
	will preserve the dominance properties of the mixed feedback amplifier. Moreover, the stability/instability
	of the closed-loop equilibria will also remain unchanged for sufficiently small perturbations. This means that
	the stable / oscillatory regimes of the mixed feedback are robust. A quantitative analysis can be developed
	through convex optimization, based on linear matrix inequalities \cite{forni2018differential,padoan2019h}.

	\section{CONCLUSIONS}
	We use dominance analysis to achieve controlled robust oscillations in closed loop, by combining positive and negative feedback.
	The design is inspired by insightful observations from biology, neuroscience, and electronics, 
	which identify in the mixed feedback a fundamental source of 
	robust and tunable oscillations. We provide a system-theoretic analysis of the mixed feedback amplifier, 
	within the simplified setting of Lure system modeling.
	Our analysis is based on dominance theory and shows how to tune feedback gain and feedback balance to control the closed-loop behavior. We later extend the analysis to larger systems. We study large parallels of positive and negative
	feedback, mimicking the structure of conductance-based models in neuroscience. We also tackle passive interconnections to control oscillations of interconnected loads. We finally discuss briefly the robustness of the mixed amplifier, a topic that will be developed in details in other publications. In fact, many directions are left to explore, starting from the problem of shaping the oscillation regime through feedback and the problem of feedback design for robustness. The problem of shaping the oscillation frequency of the mixed feedback amplifier is studied in the follow up paper \cite{che2021shaping}.
	
	\vspace*{-0.1cm}
	
	\section*{Acknowledgments}
	\vspace*{-0.1cm}
	The authors would like to thank Rodolphe Sepulchre for very insightful discussions on the 
	mixed feedback amplifier and on conductance-based models in neuroscience. These
	have been very helpful to shape our paper.


	\bibliographystyle{ieeetr}


		\appendix
	\subsection{Proof of Theorem \ref{th:0-dominance}}
	For $\lambda=0$, the problem is equivalent to the classic absolute stability problem. According to Theorem \ref{th:circle_cirteria}, the mixed feedback amplifier is $0$-dominant if the Nyquist plot of the linear system $G(s,k,\beta)$ lies to the right hand side of the line $-1$ in the complex plane. 
	Note that in the expression \eqref{eq:linear_tf}, $G(s,k,\beta)=kG(s,1,\beta)$. Hence
	the condition on the Nyquist plot of $G(s,k,\beta)$ is verified whenever $0 \leq k < \bar{k}_0$, by construction.
	
	\subsection{Proof of Theorem \ref{th:2_passive}}
	
	We will prove Theorem \ref{th:2_passive} by showing the shifted linear system $G(s-\lambda,k,\beta)$ is $2$-passive. 
	The result will then follow from the interconnection theorem \cite[Theorem 4]{forni2018differential}.
	
	The linear system $G(s,k,\beta)$ is $2$-passive if the shifted linear transfer function $G(s-\lambda,k,\beta)$ is positive real, i.e. $\angle G(s-\lambda,k,\beta)\in(-90^\circ,90^\circ)$ for all $s=j\omega$, \cite[Theorem 3.3]{miranda2018analysis}.
	For a linear system, a stable pole and an unstable zero can contribute to a $90^\circ$ phase lag,
	while an unstable pole and a stable zero can contribute to a $90^\circ$ phase lead.
	When $\beta>\beta^*$, the shifted linear system $G(s-\lambda,k,\beta)$ has two unstable poles, one stable pole and one unstable zero. 
	
\begin{figure}[!h]
		\centering
		\includegraphics[width=0.4\textwidth]{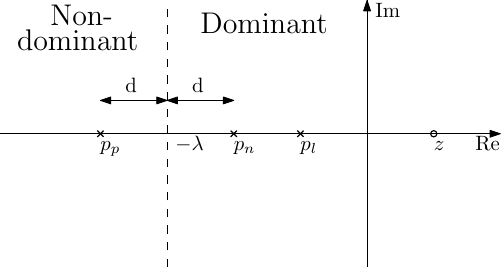}
		\caption{Setting $\lambda$ in the complex pole plane such that $G(s-\lambda,k,\beta)$ is $2$-passive}
		\label{fig:2_passive_illustration}
	\end{figure}

	\textbf{Case $\tau_n < \tau_l$:} as illustrated in Fig. \ref{fig:2_passive_illustration}, 
	take 
	$\lambda=-\frac{p_p+p_n}{2}$. As a consequence, the shifted linear system $G(s-\lambda,k,\beta)$ has a stable pole and an unstable pole of the same magnitude, meaning that their contributions to the phase of the system cancel out. The shifted system then is left with a pair of unstable pole and zero, which guarantees the positive realness of $G(s-\lambda,k,\beta)$. This property holds for any $k>0$, since $k$ does not influence the positions of poles and zeros of the open loop transfer function.
	
	\textbf{Case $\tau_l < \tau_p$ and case $\tau_p < \tau_l < \tau_n$:} same as above, taking $\lambda=-\frac{p_l+p_p}{2}$.

	{
		\begin{remark}
			The rate $\lambda$ does not necessarily need to locate exactly at the middle of the non-dominant and dominant poles, as in the proof. A small perturbation of $\lambda$ from the middle position, for example, will introduce negligible fluctuations on the phase without violating the overall positive realness of $G(s-\lambda,k,\beta)$. 
			The range of feasible $\lambda$ will be wider for larger time-scale separation between dominant and non-dominant poles.
		\end{remark}
	}
	
	\subsection{Proof of Proposition \ref{th:Interlacing2}}
	Note that $C(s)$ in \eqref{eq:Interlacing_expression} is continuous along the real axis except at the poles. Then, 
	for every $\beta\in(0,1)$ and for all $i\in(1,2,...,m-1)$ and $i\in(m+1,m+2,...,m+n-1)$,
	\begin{equation}\label{eq:interlacing_limit}
		\begin{cases*}
			\lim\limits_{s\rightarrow -1/\tau_i^+}C(s)= \infty \quad \lim\limits_{s\rightarrow -1/\tau_{i+1}^-}C(s)= -\infty\\
			\lim\limits_{s\rightarrow -1/\tau_i^+}C(s)= -\infty \quad \lim\limits_{s\rightarrow -1/\tau_{i+1}^-}C(s)= \infty \ .
		\end{cases*}
	\end{equation}
	Thus, for every $\beta\in(0,1)$, by the intermediate value theorem $C(z)$ must have at least one zero $z\in(-1/\tau_i,-1/\tau_{i+1})$, i.e. the zero $z$ is in-between the poles of $Cp(s)$ and $Cn(s)$. This continuity argument clarifies the relative position of $m+n-2$ zeros with respect to the poles of $Cp(s)$ and $Cn(s)$. Since $C(s)$ can only have a total number of $m+n-1$ zeros, we need to determine the relative position of one more zero.
	
	We will show that the last zero must locate in the interval $(-\infty,-1/\tau_1)\cup(-1/\tau_{m+n},\infty)$. 
	If any, a zero $z_0\in(-\infty,-1/\tau_1)\cup(-1/\tau_{m+n},\infty)$ guarantees that $\frac{Cp(z_0)}{Cn(z_0)}>0$, i.e. $Cp(z_0)$ and $Cn(z_0)$ have the same sign and neither of them is equal to zero. It follows that any point in 
	$z_0\in(-\infty,-1/\tau_1)\cup(-1/\tau_{m+n},\infty)$ can be made a zero of  $C(s)$ by setting $\beta=\frac{Cn(z_0)}{Cp(z0)+Cn(z_0)}\in(0,1)$. Hence we have shown that for every $z_0\in(-\infty,-1/\tau_1)\cup(-1/\tau_{m+n},\infty)$, there is a $\beta\in(0,1)$ such that $z_0$ is a zero of $C(s)$.
	
	In summary, we have shown that $C(s)$ has $m+n-2$ zeros interlaced with poles from $Cp(s)$ and $Cn(s)$ and a zero $z_0\in(-\infty,-1/\tau_1)\cup(-1/\tau_{m+n},\infty)$, which already are $m+n-1$ zeros.
	As final remark, we observe that there is a limit value of $\beta$ for which the zero $z_0$ 
	moves towards infinity, given by
	\[\beta=\lim\limits_{z_0\rightarrow\pm\infty}\frac{Cn(z_0)}{Cp(z0)+Cn(z_0)}\]
	For $m=n=1$, this limit corresponds to $\beta^*$ in Theorem \ref{th:2_passive}.

\end{document}